# Emergent Network of Josephson Junctions in a Kagome Superconductor

*Tycho J. Blom[1†], Matthijs Rog[1†], Marieke Altena[2], Andrea N. Capa Salinas[3], Stephen D. Wilson[3], Milan P. Allan[1,4,5,6], Chuan Li[2], Kaveh Lahabi[1*]*

1 Huygens-Kamerlingh Onnes Laboratory, Leiden University, 2300 RA Leiden, The Netherlands.
2 MESA+ Institute for Nanotechnology, University of Twente, 7500 AE, Enschede, The Netherlands.
3 Materials Department, University of California Santa Barbara, Santa Barbara, California 93106, USA
4 Faculty of Physics, Ludwig-Maximilians-University Munich, Munich 80799, Germany
5 Center for Nano Science (CeNS), Ludwig-Maximilians-University Munich, Munich 80799,
Germany
6 Munich Center for Quantum Science and Technology (MCQST), Ludwig-Maximilians-University Munich, Munich 80799, Germany

Materials with a Kagome lattice are intensely studied because they host exotic states that combine strong correlations and topology. Recently, critical current oscillations were observed in an unstructured flake of $CsV_3Sb_5$ . In this work, we show that the origin of these oscillations is a network of Josephson junctions intrinsic to the flake that emerges below its critical temperature. Under radio-frequency radiation, we observe quantized Shapiro steps. The sensitivity of the step height to the contact placement indicates a complex network of junctions. By performing interference studies along multiple field directions, we demonstrate that the interference effects are a result of small junctions and filamentary supercurrent flow. Upon nanostructuring the flake, prominent features of the interference pattern are preserved, illustrating the localized nature of these junctions and their stability to thermal cycles. These results pave the way for determining the exact nature of superconductivity in the $AV_3Sb_5$ family.

The kagome lattice, a network of corner-sharing triangles, provides a rich platform for the study of novel states, due to a unique interplay of geometric frustration, topology and strong electronic correlations.[1,2] Historically, research on kagome materials has focused on magnetically ordered insulators for the study of quantum spin liquids,[3] and magnetic metals that host a wide range of phenomena, including Weyl states,[4] massive Dirac points[5] and Chern magnetism.[6] However, a new class of non-magnetic superconductors $AV_3Sb_5$ (A = K, Rb, Cs) has garnered considerable interest, hosting several topological and correlated states.
The $AV_3Sb_5$ materials exhibit a transition to a charge density wave (CDW) state which breaks rotational symmetry,[7–9] followed by the appearance of unconventional superconductivity at low temperatures.[10–14] In addition, there have been reports of time-reversal symmetry (TRS) breaking,[7,8,12,15] electronic nematicity[16–18] and tunable chirality[8,9,13,19] in the CDW state, although, in all three cases, contradictory experimental results have led to active debate.[14,20] In particular, the notion of a TRS breaking CDW state has led to much interest in its origins, with several theoretical and experimental studies suggesting loop current order as a possible resolution.[15,19,21,22] Moreover, the nature of superconductivity and its interaction with this exotic CDW is a source of intrigue. The character of the superconducting state is unclear, although the general consensus seems to converge on a multiband state with a nodeless, anisotropic gap and singlet pairing.[14] Due to the competition of so many different orders, further experimental research focusing on meso- and microscopic behavior is essential for the resolution of these important questions.

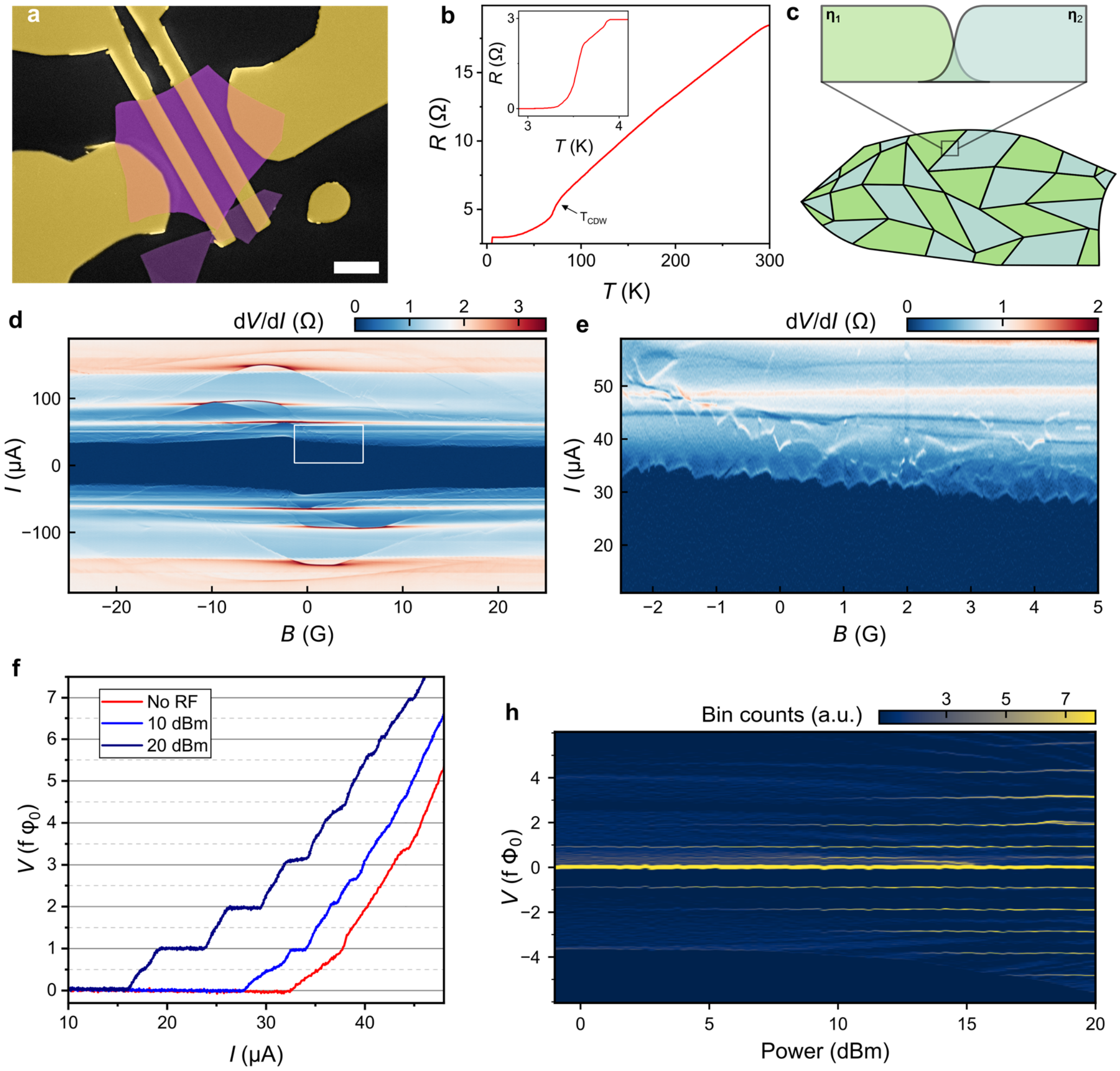

**Figure 1. (a)** Electron micrograph of a flake device. The scalebar is 5 μm. **(b)** Resistance versus temperature graph, showing the CDW transition at 73 K. Inset shows the superconducting transition. **(c)** Sketch of a superconducting domain structure as proposed in this work. The domain walls between adjacent domains function as Josephson barriers. We show a scenario with 2 colors, representing 2 types of domains, but a degenerate ground state with 3 or more distinct domains would also be possible. **(d)** Superconducting interference pattern of the device shown in (a) measured at 600 mK, showing several decaying and oscillating critical currents. **(e)** High-resolution measurement of a smaller region of the interference pattern shown in (d). **(f)** Current-voltage characteristics of the same sample measured in (d) and €, with and without 800 MHz radiation, showing Shapiro steps. Voltage scale is normalized to units of $f\phi_0$. **(h)** Power evolution of the Shapiro steps from (f). Higher steps emerge at larger applied powers.

A recent mesoscopic transport study on exfoliated flakes of $CsV_3Sb_5$ revealed a striking field-free superconducting diode effect, indicating the presence of both inversion symmetry and TRS breaking[23]. What has been largely overlooked by the community, however, is the presence of oscillating superconducting interference patterns in an external magnetic field. Initially, the oscillations were erroneously attributed to the Little-Parks effect and it was suggested that a chiral domain structure could account for these observations. Though the Fraunhofer-like shape of the envelope and the possible connection to Josephson junctions is noted, the picture presented is that edge currents form effective rings, and the associated fluxoid quantization is responsible for the oscillations.

Here, we show that this interpretation is incorrect. We present unambiguous evidence that instead, there exists a network of Josephson junctions in single, homogeneous flakes of the Kagome superconductor $CsV_3Sb_{5-x}Sn_x$ (x = 0.03-0.04) by demonstrating the presence of Shapiro steps. We argue that both the Fraunhofer envelopes and the rapid oscillations originate from Josephson interference effects, with the latter coming from parallel junctions, which form a superconducting quantum interference device (SQUID). Slight Sn-doping of $CsV_3Sb_5$ suppresses long range CDW correlations and produces a charge ordered state that is more similar to the sister compounds (K, Rb)$V_3Sb_5$.[10,11,24] Thus, together with the previous report on undoped $CsV_3Sb_5$, we believe our results generalize to the entire $AV_3Sb_5$ family.

To investigate the nature of the magnetically sensitive critical currents in $CsV_3Sb_{5-x}Sn_x$ (x = 0.03-0.04) we perform magnetotransport experiments under both DC and radio frequency (RF) current bias, both of which are well-established probes of Josephson physics. Josephson junctions can be identified in transport experiments by the appearance of Fraunhofer interference patterns, where the junction critical current ($I_c$) decays and oscillates under an applied magnetic field. However, depending on the geometry of the junction and the internal supercurrent distribution, the exact shape of the pattern need not always follow the typical Fraunhofer relation.[25,26] The radio frequency (RF) response of a device, however, allows for unambiguous identification of the AC Josephson effect. When RF radiation is introduced to the junction, discretized steps, called Shapiro steps, emerge in the current-voltage (IV) characteristics. These steps, appearing at quantized voltages $V_n = nf\phi_0$ (with $n$ an integer and $f$ the applied RF frequency) can be considered smoking-gun evidence for the existence of a Josephson junction network.

In Figure 1, we present the key data of this work: unstructured flakes showing oscillating interference patterns, and corresponding Shapiro steps. Devices were fabricated by lithographically contacting single-crystal flakes with Pt leads (see Supporting Information for more details). A micrograph of a flake device is shown in Figure 1a. Cooling down from room temperature, we see a clear signature of the CDW transition at 73 K (Figure 1b). The broad superconducting transition exhibits a nonstandard shape (Figure 1b inset).

In Figure 1d we show a superconducting interference pattern, with an applied out-of-plane (OOP) field running between -25 and 25 Gauss. Several peaks in the differential resistance are identified, all of which decay with field. These peaks are critical currents, each associated with a single layer in the Josephson junction network. A closer look reveals rapid oscillations in the lowest $I_c$, with a period of roughly 0.3 G. We stress that the area corresponding to this oscillation (~70 μm2) is very close to the area enclosed by the voltage contacts (~70 μm2), implying a substantial portion of the flake contributes to the observed interference. Our interference patterns all show a point symmetry around a field value close to zero, congruent with previous reports.[23] To show a field-free diode effect in our samples, however, would require a more careful investigation of the magnetic background.

Figure 1f shows single IV curves with and without applied RF radiation. We clearly observe the evolution of Shapiro steps at integer voltages, and faint signs of half-integer steps as well. The first steps lay cleanly on the integer lines, while the higher steps deviate slightly, with the distance between steps growing as the current increases. This is a natural consequence of the proximity of multiple critical currents, causing the RF response of the second $I_c$ to mix with the first. Nevertheless, the Shapiro steps of the lowest critical current is definitive proof of its Josephson nature, strongly implying that the rapid oscillation from Figure 1e originates from a SQUID-like interference pattern, not a Little-Parks effect.

We now demonstrate that Little-Parks oscillations are an unfit description of our interference patterns, even in the absence of Shapiro steps. The Little-Parks effect describes a magnetic field oscillation in the resistance of a superconducting loop, originating from fluxoid quantization. Flux through the loop causes a phase-induced current, which slightly decreases its critical temperature. This $T_c$ oscillation can show up in resistance measurements close to the transition. While the period of the Little-Parks effect coincides with that of SQUID oscillations, the resemblance is otherwise superficial. In particular, Little-Parks oscillations can only appear near the superconducting transition.[27] The interference patterns in Figure 1 were taken at 600 mK, well below the superconducting transition, so they cannot be a manifestation of the Little-Parks effect, even if there would be a ring-like geometry in the sample.

Given the absence of obvious conventional weak links, a natural next line of inquiry would involve the origin and spatial distribution of these junctions. While a complete resolution of these questions lies beyond the scope of this work, we can draw some conclusions from our transport data. Based on the number of critical currents, and the frequent appearance of SQUID and Fraunhofer oscillations, we infer that there is a network of junctions in this material. The presence of SQUID patterns also suggests filamentary currents, since a homogeneous current distribution through the flake could not result in two separate parallel junctions. The spontaneous emergence of junctions bears resemblance to the intrinsic junctions found in layered superconductors with a weak inter-layer coupling.[28–30] In our devices, however, the direction of transport is predominantly along the Kagome layers (the a-b plane), as a natural result of exfoliation and our contacting methodology. In addition, the response of our junctions to an OOP field shows that the junctions transport direction cannot point along the c-axis.

A possible scenario that could account for in-plane junctions, which has been previously suggested,[23] is a superconducting domain structure. A multicomponent order parameter, such as a chiral pairing, could result in a degenerate superconducting ground state. The resulting domains, and in particular the domain walls between them, have been reported to form stable Josephson junctions.[27] Figure 1c depicts a schematic overview of this idea. We make no claims about possible chirality, nor that there are necessarily exactly 2 distinct domains. If the intrinsic junctions all have the same critical current density, the origin of the large difference in measured $I_c$ could be different domain wall dimensions being encountered along the inhomogeneous current paths.

Some reports have shown that CDWs themselves can also respond to applied RF radiation[31–33] or acoustic vibration,[34,35] in a mechanism analogous to the AC Josephson effect, to produce Shapiro-like steps. The resulting steps are quantized, but result in sharp peaks in the differential resistance, instead of the minima associated with regular Shapiro steps. The interplay between the CDW and the AC Josephson effect can result in some non-trivial signatures,[36–38] but does not distract from the existence of the junctions in the first place. More importantly, these dynamics only present themselves in an incommensurate CDW whereas the CDWs in the $AV_3Sb_5$ family are known to be commensurate,[7,9,11] with incommensurate charge correlations arising only in highly Sn-doped $CsV_3Sb_{5-x}Sn_x$ (x = 0.15).[39]

We note the similarity of our results to another recent report on $NbSe_2$, another system with superconductivity and CDW order.[38] Here, Shapiro steps are also observed in a flake with no apparent weak link. The authors state that in this device, a phase-slip line (PSL) forms an effective junction. PSLs, lines of kinematic vortices, are known to act as weak links in wide superconducting strips.[40] Our junctions, however, lack a number of characteristic features of PSLs, including a uniform distribution of junctions along the sample, equal excess current of all resistive branches, and differential resistance of a branch proportional its number (i.e., the second branch, above the second $I_c$, having twice the differential resistance as the first).[40,41] We thus conclude that PSLs cannot account for our junctions.

In several samples, we observe discrete voltage steps in the RF response that are not integer units of $f\phi_0$. However, these "non-integer" steps always scale linearly with frequency and grow with applied RF power, which suggests that they indeed originate from the Shapiro effect. We now demonstrate that this is indeed the case, and that all these samples exhibit the integer Shapiro effect, with the non-integer steps a measurement artefact.

Fractional, or subharmonic, Shapiro steps are known to arise in Josephson systems with a multivalued or anharmonic current-phase relation.[42–44] Fractional Shapiro steps always occur alongside integer steps, and the simpler fractions (1/2, 1/3, 2/3, etc.) always produce more pronounced steps than smaller or more complex fractions. Furthermore, multiple different fractions are expected, not just integer multiples of one specific fraction. Our Shapiro steps do not resemble this typical fractional behavior at all, instead appearing more like scaled down integer steps.

The key insight comes from the fact that the spectrum of steps changes drastically when switching the current and voltage probes. Figure 2a shows a spectrum of non-integer Shapiro steps, measured at 800 MHz, in the conventional four-probe configuration, with the voltage probes on the inside. When we instead apply current through the inner probes, and measure voltage over the outside, as shown in Figure 2b, the Shapiro steps become fully integer (see Supporting Information for additional data).

In a normal junction device, switching the probe configuration should not result in any difference in the Shapiro response, as the same part of the device is being probed. However, in a randomly distributed network of junctions, the likelihood is high that at least one junction is partly covered by a voltage contact. As sketched in Figure 2e, when a junction is partly covered by one of the voltage leads, it picks up the potential of both sides of the junction, depending on the exact area on each side. The measured voltage will thus be only a part of the entire voltage drop across the junction (which is a normal integer voltage step). This scenario is likely to occur, since the contacts are patterned without any knowledge of the spatial distribution of junctions.

When inverting the probes, we measure voltage on the outside of the flake, which means the full potential drop of a junction is always taken into account, regardless of which junction has the lowest $I_c$. Consequently, exclusively integer steps are observed when the probes are inverted. We therefore conclude that the junctions themselves harbor no exotic fractional states, but rather that the observed non-integer steps are a measurement artefact.

Inverting the contacts also affects the interference pattern, as shown in Figure 2c and Figure 2d. Here many features of the pattern coincide, indicating that we primarily probe the same area of the flake. The difference can be explained precisely by the area covered by the inner contacts. When using these leads as voltage probes, the area will always contribute to the measured voltage, but when applying current, the current will avoid junctions to lower the Josephson energy of the network.

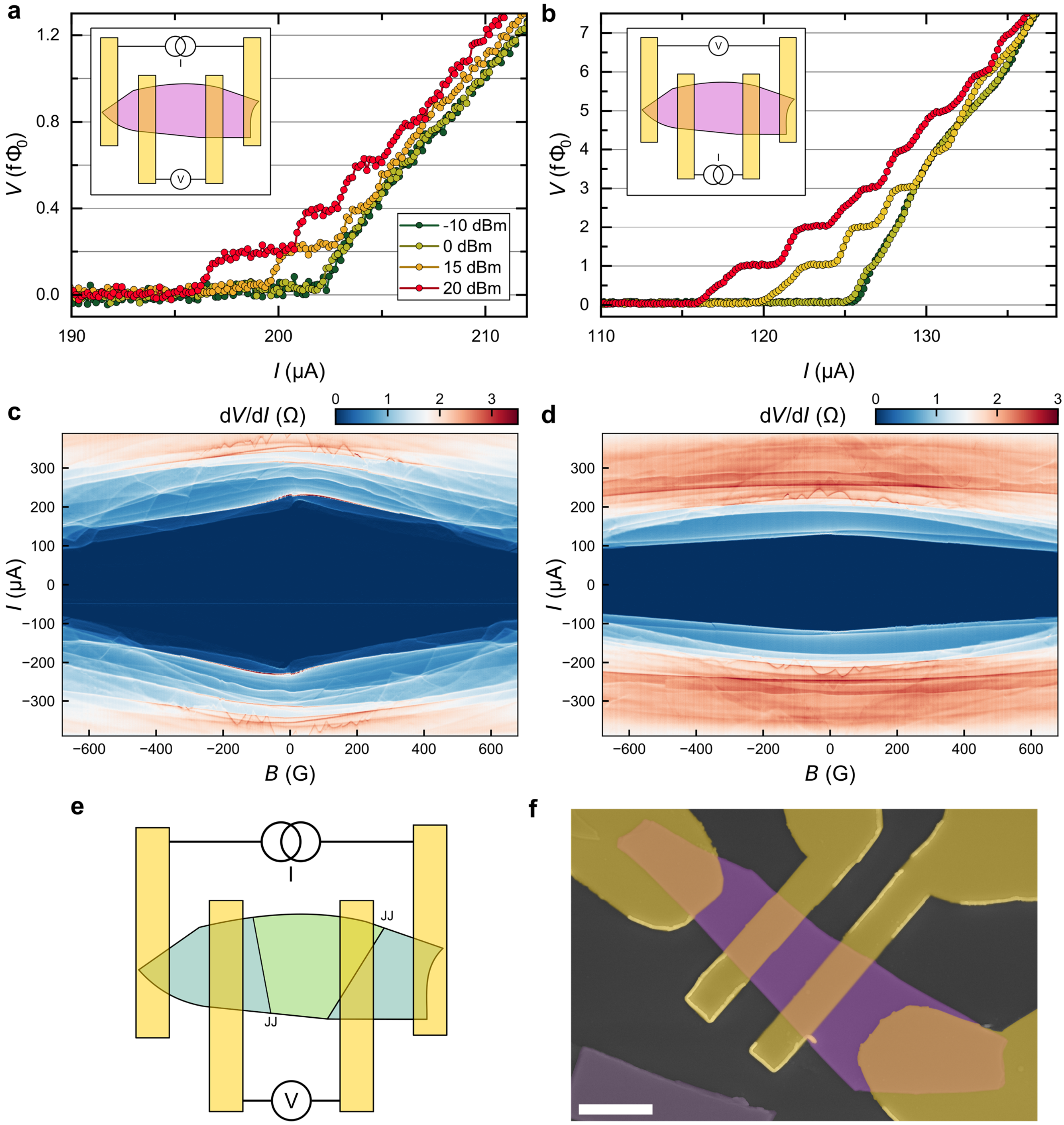

**Figure 2.** **(a)** Shapiro response of flake shown in (f) to irradiation of 800 MHz frequency, measured at 600 mK. The normalized voltage scale shows steps at non-integer multiples of $f\phi_0$. Inset shows the four-probe measurement configuration. A small slope and offset has been subtracted. **(b)** The same measurement, but now the current and voltage leads are switched, as shown in the inset. The Shapiro steps occur at integers. **(c)** and **(d)** Superconducting interference patterns for the configurations shown in (a) and (b), respectively. **(e)** Sketch of a flake with 2 emergent junctions, revealing the origin of the non-integer Shapiro steps. **(f)** Electron micrograph of the flake. The scalebar is 5 μm.

The nature and nanoscale structure of the intrinsic Josephson junction network in these flakes is further elucidated by their interference patterns for different, orthogonal orientations of the external field.[26,45] Figure 3 shows such an experiment on a large flake, where the transport characteristics again distinguish a cascade of critical currents, implying a complex network of junctions in this sample.

Figure 3a and Figure 3b, respectively, show supercurrent interference patterns with the field oriented fully OOP, and fully in-plane (IP). In both cases, four major critical currents, labeled by arrows, can be distinguished. At zero field, these critical currents behave similarly, yet their response to the external field is completely different.

The properties of the interference patterns can be understood within the framework of Josephson junction networks. Again, parallel junctions exhibit rapid SQUID oscillations at a period proportional to the loop area. In Figure 3a, critical current 2 shows such rapid oscillations with a period, 1.2 mT, corresponding to a loop area of 1.7 μm$^2$. The fact that such small loops form a dominant contribution to the transport in an otherwise large (>100 μm$^2$) flake, indicates highly inhomogeneous current flow. Such rapid oscillations are absent from critical currents 1, 3 and 4. We thus identify these critical currents as corresponding to single junctions in the Josephson network, which only show Fraunhofer-like decay when a field is applied. In the OOP interference pattern, critical currents 1 and 3 clearly inhabit this category while the envelop of critical current 2 also indicates Fraunhofer interference inside the loop's constituent junctions.

Crucially, Fraunhofer interference remains visible when orienting the magnetic field IP, as long as the direction of field is perpendicular to the current direction. However, the effective area is reduced: the junction width is now replaced by the junction thickness. This is in sharp contrast with SQUID interference, which must completely disappear, as the projected area of any loop in the ab-plane goes to zero for an IP field.

Figure 3b shows that the IP interference effect exactly corroborates these expectations. The rapid oscillations of critical current 2 disappear, as the projected area of the SQUID loop goes to zero. For all critical currents, the Fraunhofer envelope is still visible, and decays with the same shape as its OOP counterpart, but over a larger field range.

Accounting for the angle between the (global) current direction and the IP field (see Figure 3d) and taking the full thickness of the flake (101 nm) as the junction thickness, various important quantities can be inferred, most easily for critical current 1. We estimate a junction width of 1.2 μm and a junction length of 1.3 μm. Such small junction widths compared to the device width (>10 μm) are congruent with the critical currents observed: the corresponding critical current densities are of the same order of magnitude as those observed for other metallic junctions.[27,46,47]

To further study the structure of the junction network, we use focused ion beam (FIB) milling to cut a bar of 3.2 μm width, shown in Figure 3d, out of the same flake, and repeat the experiment. Figures 3c and 3f show the resistance-temperature characteristic of the sample before and after nanostructuring. Above $T_C$, the resistance increase precisely matches the reduction in device width (~3.7x). However, the height of the shoulder is unresponsive to the nanostructuring, with its resistance slightly decreasing. This indicates that this is not a normal metal resistance, but a property of the forming Josephson network.

Figure 3e shows the IP interference pattern of this flake, taken at the exact same temperature and field angle as the pre-FIB pattern. The two patterns share many similarities. Most strikingly, despite the confinement of the current, critical current 1 appears again, with the exact same critical current and interference pattern. Both above and below critical current 1, new critical currents appear.

These observations are fully consistent with the Josephson network picture. After nanostructuring, the current is forced to travel a different path. However, where-ever the new path overlaps with the old, the same junctions are included in the network, leading to identical interference features. The width reduction has no impact on these critical currents, because in filamentary networks the current is never homogeneously distributed over the entire width of the bar.

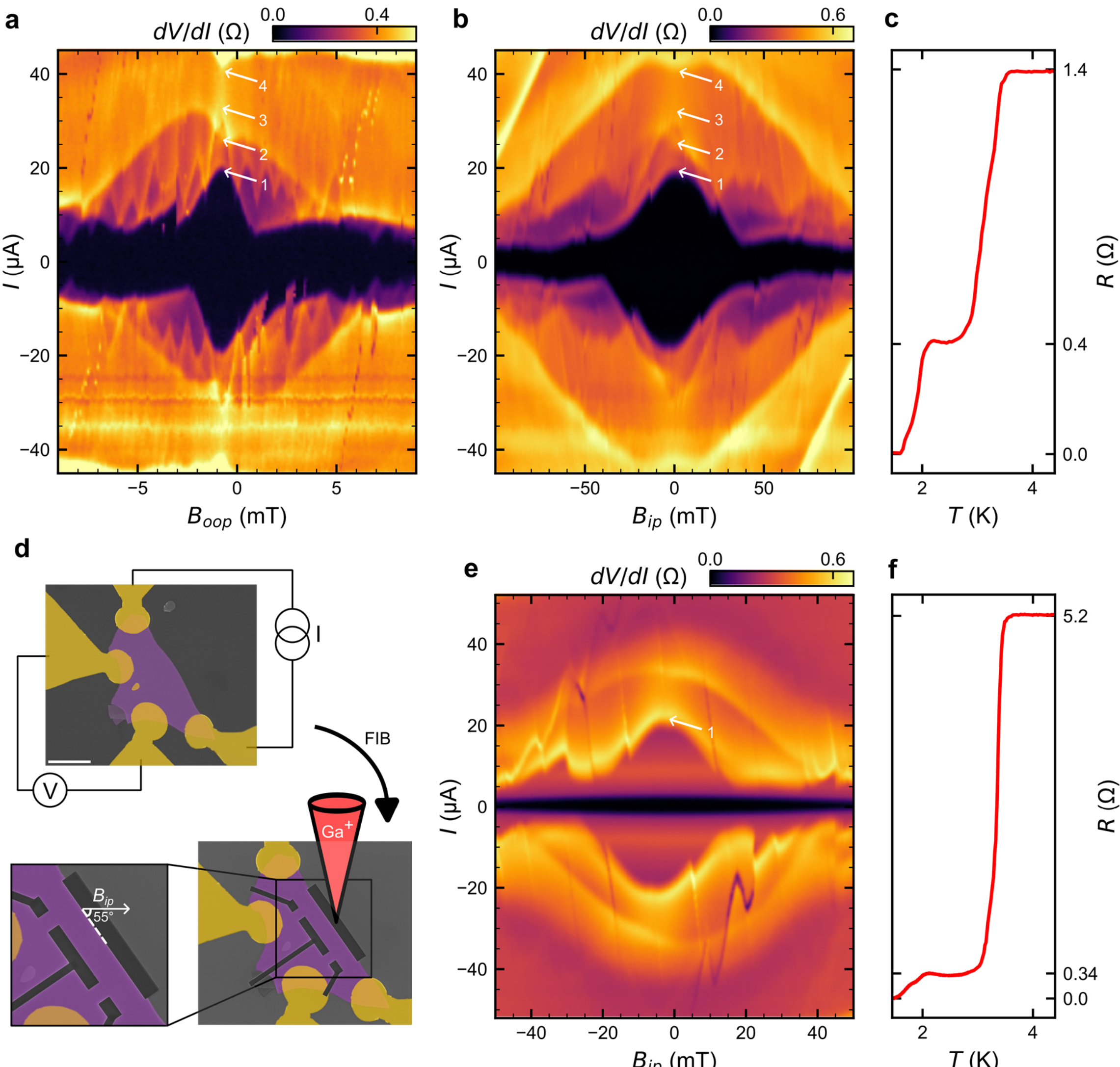

**Figure 3.** Supercurrent interference study using current confinement and various components of the magnetic field. a OOP interference pattern. **(a)** OOP interference pattern. **(b)** IP interference pattern. **(c)** Resistance-temperature characteristic of the device used in the interference patterns. The device has a highly inhomogeneous superconducting transition. **(d)** Image of the device studied. Using FIB milling, a bar of width 3.2 µm is structured, strongly confining the currents. **(e)** IP interference pattern taken after nanostructuring. New critical currents appear, but critical current 1 is completely unchanged compared to the pattern in (b), indicating filamentary current flow partially undisturbed by the structuring. **(f)** Resistance-temperature characteristic of the device after FIB. Above the transition, the new resistance is fully consistent with the new device dimensions. The wide transition, however, is not strongly impacted by the new geometry, implying this resistance is not metallic, but an attribute of the forming Josephson network.

To summarize, we present unequivocal evidence for the presence of an emergent network of Josephson junctions in the Kagome superconductor $CsV_3Sb_{5-x}Sn_x$. Our results further highlight the unusual properties of Kagome materials and show that it is also determined by the intriguing physics of mesoscopic Josephson networks, which lead to Shapiro steps as well as Fraunhofer and SQUID oscillations. In the future, further nanostructuring could be combined with local sensing to further elucidate the complex nanoscale structure of the intrinsic junction network. Furthermore, the approaches presented here can be used to demonstrate the Josephson effect in other devices where critical current oscillations far below $T_C$ are attributed to a Little-Parks effect, such as those in $MoTe_2$.[48]

During the preparation of this manuscript, we became aware similar results in $CsV_3Sb_5$ simultaneously obtained by two other groups.[49,50]

**Corresponding Author**

* Kaveh Lahabi, Huygens-Kamerlingh Onnes Laboratory, Leiden University, P.O. Box 9504, 2300 RA Leiden, The Netherlands; Email: lahabi@physics.leidenuniv.

**Author Contributions**

The manuscript was written through contributions of all authors. All authors have given approval to the final version of the manuscript. † These authors contributed equally.

**Funding Sources**

This work was supported by the Dutch Research Council, NWO, as part of the “The Full Picture: Novel Microscopy with Smart Quantum Probes” (Project: VI.Veni.212.302) and “Three seeds for the quantum/nano-revolution” (Project: NWA 1418.22.001). It was also partially supported by “ERC CoG pairNoise”. SDW acknowledges support from the U.S. Department of Energy (DOE), Office of Basic Energy Sciences, Division of Materials Sciences and Engineering under grant no.DE-SC0020305.

**Acknowledgements**

The authors would like to thank Daan Wielens, Titus Neupert and Mark Fischer for helpful discussions.